\documentclass[structabstract]{aa}
\usepackage{natbib}
\usepackage{txfonts}
\usepackage{graphicx}
\titlerunning{Projection factor and limb darkening}
\authorrunning{H.~R. Neilson}

\begin{document}

\title{Cepheid limb darkening, angular diameter corrections, and projection factor from static spherical model stellar atmospheres}

\author{Hilding R. Neilson\inst{1} \and Nicolas Nardetto\inst{2} \and Chow-Choong Ngeow\inst{3} \and Pascal Fouqu\'{e}\inst{4} \and Jesper Storm\inst{5}}
\institute{
     Argelander-Institut f\"{u}r Astronomie, Universit\"{a}t Bonn, Auf dem H\"{u}gel 71, D-53121 Bonn, Germany \\
     \email{hneilson@astro.uni-bonn.de} 
     \and
     Laboratoire Lagrange, UMR7293, UNS/CNRS/OCA , 06300 Nice, France
     \and
     Graduate Institute of Astronomy, National Central University, Jhongli City, 32001, Taiwan 
     \and
     IRAP, Universit\'e de Toulouse, CNRS, 14 avenue Edouard Belin, 31400 Toulouse, France
     \and
     Leibniz-Institut f\"{u}r Astrophysik Potsdam (AIP), An der Sternwarte 16, 14482, Potsdam, Germany
    }

\date{}

\begin{abstract}
{One challenge for measuring the Hubble constant using Classical Cepheids is the calibration of the Leavitt Law or period-luminosity relationship. The Baade-Wesselink method for distance determination to Cepheids relies on the ratio of the measured radial velocity and pulsation velocity, the so-called projection factor and the ability to measure the stellar angular diameters.}
{We use spherically-symmetric model stellar atmospheres to explore the dependence of the p-factor and angular diameter corrections as a function of pulsation period.}
{Intensity profiles are computed from a grid of plane-parallel and spherically-symmetric model stellar atmospheres using the \textsc{SAtlas} code. Projection factors and angular diameter corrections are determined from these intensity profiles and compared to previous results.}
{Our predicted geometric period-projection factor relation including previously published state-of-the-art hydrodynamical predictions is not with recent observational constraints. We suggest a number of potential resolutions to this discrepancy. The model atmosphere geometry also affects predictions for angular diameter corrections used to interpret interferometric observations, suggesting corrections used in the past underestimated Cepheid angular diameters by $3-5\%$.}
{While spherically-symmetric hydrostatic model atmospheres cannot resolve differences between projection factors from theory and observations, they do help constrain underlying physics that must be included, including chromospheres and mass loss.  The models also predict more physically-based limb-darkening corrections for interferometric observations.}
\end{abstract}
\maketitle

\section{Introduction}
Classical Cepheids form an important rung on the cosmic distance ladder because they follow a period-luminosity relation, also called the Leavitt Law \citep{Leavitt1908}.  As such, Cepheids are used to measure distances to other galaxies \citep[e.g.][]{Piet2010}, and cosmological parameters \citep{Freedman2001, Riess2009, Riess2011}, but the Leavitt Law must be calibrated using Galactic \citep[e.g.][]{Turner2010} and Large Magellanic Cloud (LMC) Cepheids \citep[e.g.][]{Ngeow2008,Ngeow2009}.

The Leavitt Law is calibrated by measuring Cepheid distances using  numerous of techniques, such as parallax, cluster membership, and the Baade-Wesselink method. The Leavitt Law has been derived by assuming all LMC Cepheids are at the same distance, where the distance to the LMC is derived using a number of methods \citep[e.g.][]{Gieren2005, Clement2008, Bonanos2011}.    Parallax has been measured for a number of Galactic Cepheids \citep{Leewen2007, Benedict2007}, but this method is limited by the precision of Hipparcos and the Hubble Space Telescope.  Only a small number of Cepheids have precise parallaxes measured, and this will only change when the GAIA satellite is launched \citep{Windmark2011}.  Likewise, \cite{Turner2010} measured distances to $24$ Cepheids belonging to clusters and groups.  On the other hand, the Baade-Wesselink method can be used for Galactic and LMC Cepheids \citep{Kervella2004, Gieren2005}, allowing for a much greater sample size.

The Baade-Wesselink method \citep{Baade1926, Wesselink1946} measures the distance to a Cepheid by measuring the change of angular diameter and the actual change of radius as a function of phase.  These measurements yield the mean radius and mean angular diameter, hence the distance.  The change of angular diameter can be determined using a number of methods.  One method is from interferometric observations, where the angular diameter can be directly resolved \citep{Kervella2004, Kervella2006, Merand2006}. The method uses angular diameter corrections for limb darkening when the observations have limited resolution.  A second method is using the infrared surface brightness technique \citep{Barnes1976, Gieren1989}.  This method uses a correlation between the angular diameter at any phase and the star's $V$-band flux and color. This method is derived from the relation $F_{\rm{Bol}} \propto \theta^2 T_{\rm{eff}}^4$, where $\theta$ is the angular diameter, $F_{\rm{Bol}}$ is the bolometric flux and $T_{\rm{eff}}$ is the effective temperature.  The effective temperature is correlated to the color based on observations of other stars or model stellar atmospheres, while the bolometric flux is replaced by the $V$-band flux and the bolometric correction.  This method assumes the radius is defined at a specific location in the stellar photosphere, i.e. at an optical depth $\tau = 2/3$.

The change of radius is $\Delta R = \int v_{\rm{puls}} dt$, where $v_{\rm{puls}}$ is the pulsation velocity. One cannot directly measure the pulsation velocity but instead  only the radial velocity.  The measured radial velocity is the projection of the pulsation velocity weighted by the intensity at each point on the stellar disk, i.e.
\begin{equation}\label{e1}
v_{\rm{rad}} = \frac{\int v_{\rm{puls}}I(\mu)  \mu^2 d\mu}{\int I(\mu) \mu d\mu} = v_{\rm{puls}} \frac{K}{\mathcal{H}} = \frac{1}{p}v_{\rm{puls}}.
\end{equation}
The intensity, $I(\mu)$, is given as a function of the cosine of the angular distance between a point on the stellar disk and the center of the disk, defined as $\mu$.  The quantities $\mathcal{H}$ and $K$ are the second and third moment of intensity, respectively \citep{Mihalas1978}. \cite{Getting1935} defined the ratio $\mathcal{H}/K$ to be the projection factor or p-factor, and it is a function of the intensity profile of a Cepheid at any phase. \cite{Getting1935} determined a projection factor $p=1.41$ based on an assumed limb-darkening profile. 

This p-factor was the accepted value until \cite{Parsons1972} measured it by fitting observed spectral lines using line bisectors to compare with theoretical spectra.  By measuring the line center and the doppler shift from the line center of a static atmosphere, one can determine the projection factor. \cite{Parsons1972} measured $p = 1.31$. \cite{Hindsley1986} used a similar method to measure how the p-factor varies as a function of pulsation period $p =-0.03\log P + 1.39$.  Similarly, \cite{Sabbey1995} using hydrodynamic models to measure the p-factor as a function of pulsation phase, found $p = 1.3$ to $p = 1.6$. Recent theoretical measurements based on spectral line profiles from hydrodynamical spherically-symmetric model stellar atmospheres, with HARPS spectroscopic measurements, \cite{Nardetto2004, Nardetto2007} yielded both a geometric (their Eq. 5) and a dynamic Pp relation,  assuming that radial velocities are determined using the first-moment method. On the other hand, \cite{Nardetto2009} measured a steeper slope for the dynamic relation from spectra when radial velocities are determined using a cross-correlation technique,
$p = - 0.08(\pm0.05)\log P+ 1.31(\pm0.06) $, hence referred to as the cross-correlated hydrodynamic Pp relation. The cross-correlation method for measuring radial velocities lead to a Pp relation that is about $5\%$ smaller than the Pp relation based on the first moment method. This suggests that the p-factor is a function of wavelength and is also sensitive to the technique used to measure radial velocities.
It should be noted that \cite{Gray2007} proposed a toy model for fitting the pulsation velocity to spectral line profiles directly to avoid the need of a p-factor, though this result is correlated to limb-darkening relations employed by the authors.  When their predictions of pulsation and radial velocity are converted to a p-factor for the same observations as previous works, they find larger values of the p-factor, by about $5\%$.

On the other hand, the projection factor has also been measured by using the infrared surface brightness technique to observe Cepheids with known distances. \cite{Gieren2005} required the period-projection factor relation to be $p =  - 0.15\log P+ 1.58$ to measure a distance to the LMC that is independent of Cepheid pulsation period.  \cite{Storm2011b} found similar results for LMC Cepheids, also using the infrared surface brightness technique, for different observations.
Furthermore, \cite{Storm2011a} used observations of Galactic Cepheids with known HST parallaxes to verify the steeper period dependence of the projection factor; we refer to this type of analysis as the LMC-HST Pp relation and p-factor.  \cite{Merand2005} and \cite{Groenewegen2007} measured the p-factor for Galactic Cepheids with known distances using interferometric and spectroscopic observations, both authors measured $p = 1.27$. One might argue that this method does not measure the actual value of the p-factor, but measures a ``fudge'' p-factor that corrects for uncertainties in the distance measurement. Another challenge is that angular diameter measurements imply the radius is defined at a layer in the photosphere that is the same as the layer that defines radial velocity observations; this layer may differ.

Observational and theoretical measurements of the p-factor disagree, leading to a significant problem if one wishes to use the Baade-Wesselink method to calibrate the Leavitt Law as well as understanding the metallicity dependence of the Leavitt Law. The uncertainty of the p-factor is one of the greatest sources of uncertainty for calibrating the Leavitt Law. The differences between observational and theoretical measurements are probes of unknown and uncertain physical effects in Cepheids.  In this work, we consider one possible source of the differences, i.e. stellar limb darkening, which is one of the most important contributors for determining the p-factor.  Specifically, we explore how stellar limb darkening from hydrostatic model stellar atmospheres, described in the next section,  varies as a function of fundamental parameters, themselves defined as functions of pulsation period from the literature. We explore how these intensity structures depend on the assumed geometry of the model atmospheres in Sect.~3. In Sect.~4, we compute the period projection factor (Pp) relation from the predicted intensity profiles and compare to previous results, and hypothesize why the LMC-HST and cross-correlated hydrodynamic Pp relations differ in Sect.~5.  We summarize the analysis in Sect.~6.

\section{Model stellar atmospheres}
Model stellar atmospheres are computed for both plane-parallel and spherically-symmetric geometries. We compute spherically-symmetric models using the \textsc{SAtlas} code \citep{Lester2008}.  This code is based on the \cite{Kurucz1979} \textsc{Atlas} code but updated for geometry in Fortran 90.  Plane-parallel model atmospheres are computed using our Fortran 90 version of the code.  These codes assume the stellar atmosphere is in local thermodynamic equilibrium and hydrostatic equilibrium.  Radiation transfer is computed as  a function of wavelength and intensity profiles are computed for 1000 $\mu$-points for increased precision.  \cite{Kurucz1979} models predict intensities for 17 or less $\mu$-points, while \cite{Claret2008} and \cite{Claret2011} compute limb-darkening laws from intensity profiles containing 100 $\mu$-points. The large number of $\mu$-points are necessary to adequately compare predictions and observed intensity profiles \citep[e.g.][]{Neilson2008, Neilson2011}.

The model stellar atmospheres are described by various fundamental parameters, a plane-parallel model by specifying the effective temperature and gravity, while a spherically-symmetric model requires an additional parameter such as mass, radius, or luminosity. We compute model atmospheres as a function of period.  The gravity of a Cepheid is given by the relation \citep{Kovacs2000} 
\begin{equation}
\log g = -1.21\log P + 2.62,
\end{equation}
while the radius is given by the period-radius relation determined by \cite{Gieren1999},
\begin{equation}
\log (R/R_\odot) = 0.680\log P + 1.146.
\end{equation}
The bolometric luminosity is determined by the bolometric Leavitt Law \citep{Turner2010},
\begin{equation}
\log (L/L_\odot) = 1.148\log P + 2.415,
\end{equation}
and by combining the bolometric luminosity and radius, the effective temperature is given as a function of period.

We note that using these relations implicitly assumes a period-mass relation.  This is contentious because Cepheid masses determined by different methods yield different values, i.e. the Cepheid mass discrepancy \citep{Keller2008,Neilson2011a}.  However, this does not affect the results, as the atmospheric structure is insensitive to small ($10$ - $20\%$) changes of stellar mass.

For this work, model stellar atmospheres are computed as a function of pulsation period for both geometries, assuming a range of pulsation periods from $\log P = 0.4$ - $2$ in steps of $0.1$, covering the known period range for Galactic Cepheids. For each model atmosphere, intensity profiles are computed for the $BVRIH,$ and $K$-bands, though for most of this work, discussion is restricted to $V$ and $K$-band profiles. It is these two wavebands that are used for interferometric observations \citep{Kervella2004, Mourard2010} as well as for the infrared surface brightness technique \citep{Gieren2005, Storm2011a, Storm2011b}.
\begin{figure*}[t]
\begin{center}
\includegraphics[width=0.5\textwidth]{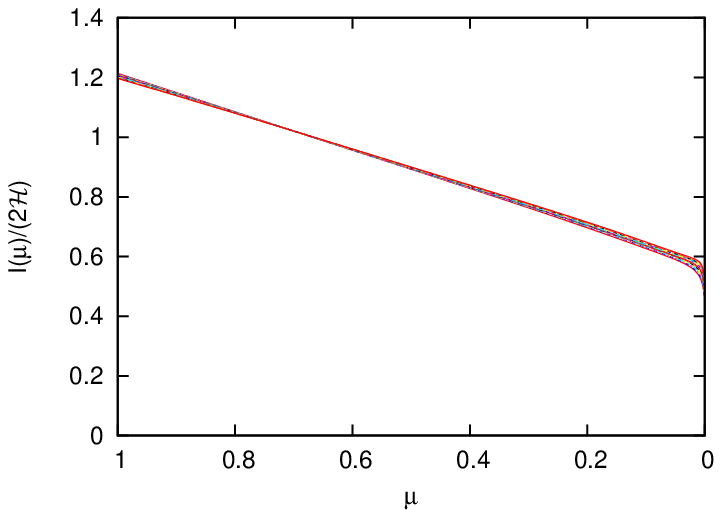}\includegraphics[width=0.5\textwidth]{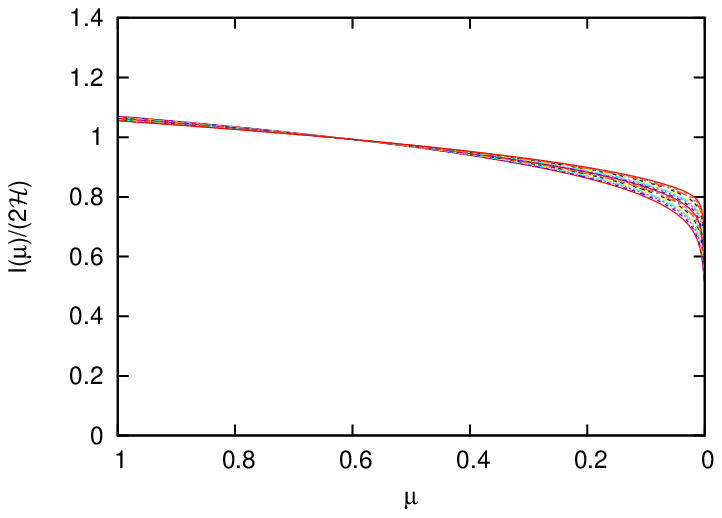}
\includegraphics[width=0.5\textwidth]{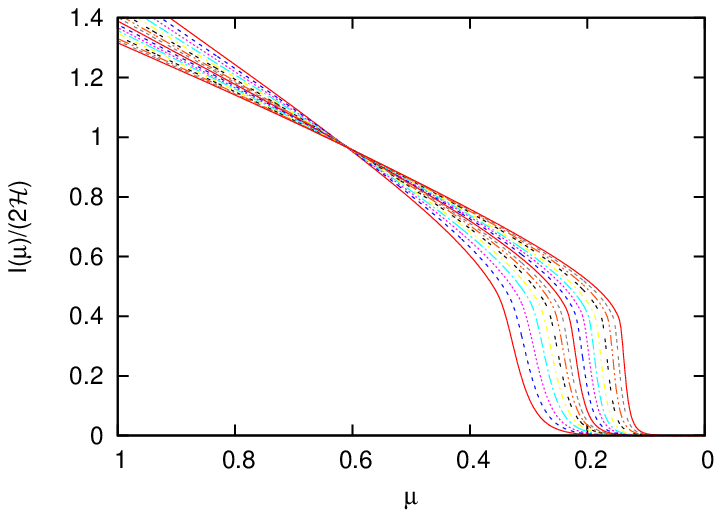}\includegraphics[width=0.5\textwidth]{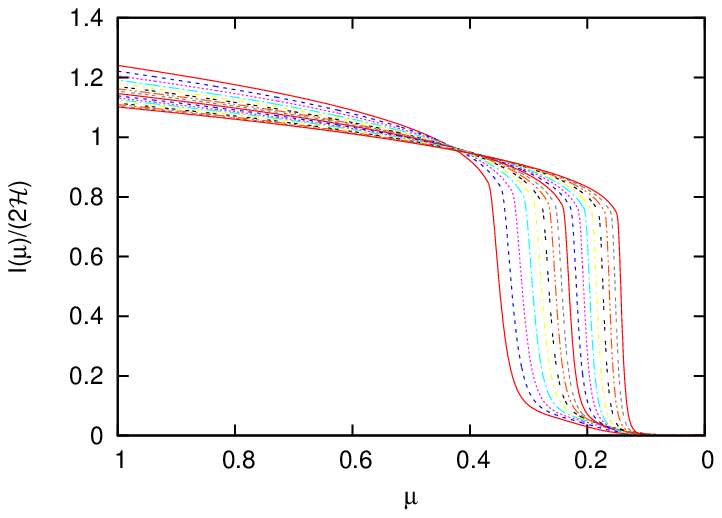}
\includegraphics[width=0.5\textwidth]{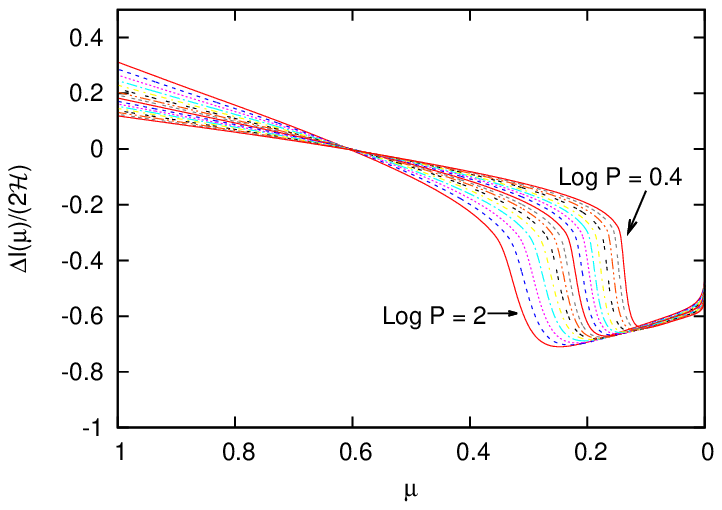}\includegraphics[width=0.5\textwidth]{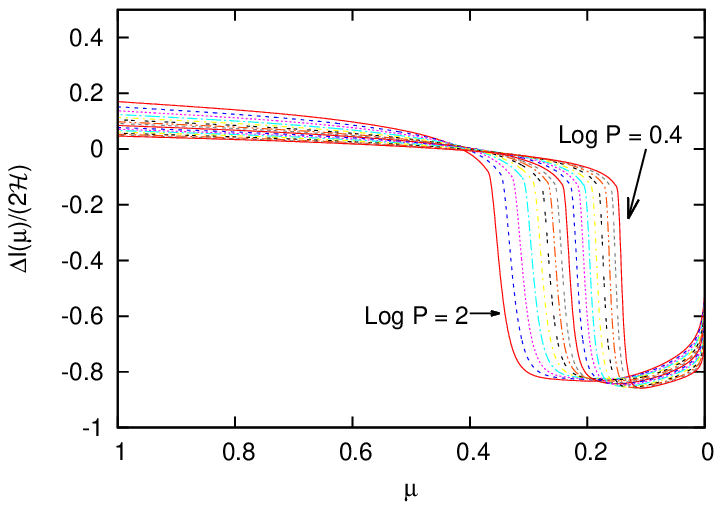}
\end{center}
\caption{Predicted intensity profiles normalized with respect to the Eddington flux for plane-parallel model stellar atmosphere (top) and spherically-symmetric model atmospheres (middle) in $V$-band (left) and $K$-band (right) wavebands.  The bottom panels show the difference between intensity profiles computed from spherically-symmetric and plane-parallel model atmospheres. Spherical models predict much smaller intensities near the stellar limb relative to the plane-parallel models. For spherical models, intensity profiles for short-period Cepheids have smaller central intensities and sharp declines at smallest value of $\mu$. Each line represents a step of $\Delta \log P = 0.1$ from $\log P =0.4$ - $2$. } \label{f1}
\end{figure*}

\section{Intensity profiles}
We explore how intensity profiles vary as a function of pulsation period. The $V$- and $K$-band intensity profiles, normalized with respect to the stellar flux, are shown in Fig.~\ref{f1} for both geometries as a function of pulsation period.  Intensity profiles from plane-parallel model stellar atmospheres do not vary much, especially in the $V$-band.  Conversely, intensity profiles from spherically-symmetric model stellar atmospheres vary significantly because of the atmospheric extension of the model atmospheres that increases as a function of pulsation period.  This atmospheric extension cannot be modelled by plane-parallel atmosphere codes.

Plane-parallel intensity profiles tend to have smaller central intensities than spherically-symmetric model intensity profiles. A model described by some pulsation period will have a specific value of luminosity and flux that must be conserved for any intensity profile regardless of geometry, i.e. for the same stellar flux,
\begin{equation}
\int I_{\rm{pp}}(\mu)\mu d\mu = \int I_{\rm{sph}}(\mu)\mu d\mu.
\end{equation}
Because spherically-symmetric model atmospheres account for atmospheric extension then the intensity towards the stellar limb for a spherical model is smaller than that for a plane-parallel model atmosphere, i.e. $I_{\rm{sph}}(\mu \rightarrow 0) \le I_{\rm{pp}}(\mu \rightarrow 0)$.  Thus, to conserve emergent stellar flux then the central intensity for the given spherical model must be greater than the central intensity for plane-parallel model atmosphere described by the same fundamental parameters. This is equivalent to noting that in spherically-symmetric models the flux diverges as a function of optical depth, leading to a steeper temperature gradient, hence a sharper decrease of intensity as a function of $\mu$. The mean intensity, $J \equiv \int I(\mu) d\mu$, along with the central intensity, increases as well. This difference is clear in the bottom panels of Fig.~\ref{f1}, where we show the difference between $V$- and $K$-band flux-normalized intensity profiles from spherically-symmetric (sph) and plane-parallel (pp) model atmospheres, $\Delta I = I_{\rm{sph}}(\mu) - I_{\rm{pp}}(\mu)$ as a function of pulsation period.  The difference is greatest near the stellar limb, depending on the atmospheric extension of the spherical model atmospheres, whereas towards the center of the stellar disk the spherically-symmetric intensity profiles are brighter than the plane-parallel model atmospheres. 

We can take this a step further and define the variable 
\begin{equation}
\sigma_I \equiv \frac{|J_{\rm{sph}} - J_{\rm{pp}}|}{J_{\rm{pp}}}.
\end{equation}
Unlike the stellar flux, the mean intensity is not required to be the same for stellar atmospheres of differing geometry, but described by the same fundamental parameters. Therefore, $\sigma_I$ is a relative measure of the model atmosphere's deviation from plane-parallel geometry, i.e. the amount of atmospheric extension. We plot this fractional difference between the mean intensity for each geometry, $\sigma_I$, as a function of pulsation period for $V$- and $K$-band intensities.
At short periods, $\log P \approx 0.4$, the difference between the mean intensities for spherically-symmetric and plane-parallel model atmospheres is approximately $20\%$, reflecting the amount of atmospheric extension for a stellar atmosphere with $\log g \approx 2.1$. At $\log P = 2$, the difference is almost $50\%$.  
\begin{figure}[t]
\begin{center}
\includegraphics[width=0.5\textwidth]{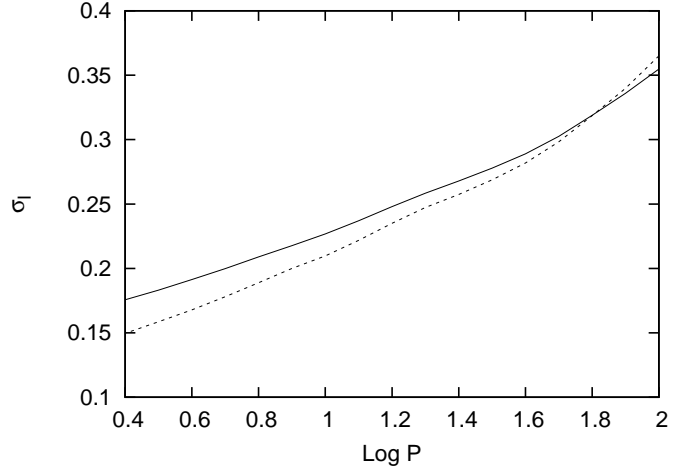}
\end{center}
\caption{The relative total intensity difference between spherical and plane-parallel model intensity profiles as a function of pulsation period, for the $V$-band (solid line) and $K$-band (dashed line) intensity profiles.} \label{f3}
\end{figure}

The results of Fig.~\ref{f3} suggest that Cepheids cannot be physically represented by plane-parallel model stellar atmospheres and plane-parallel radiative transfer.  \cite{Marengo2002, Marengo2003} computed hydrodynamic stellar atmospheres using the \cite{Sasselov1992} code, radiative transfer was computed using the plane-parallel \textsc{Atlas} code \citep{Kurucz1979}. We showed in Fig.~\ref{f1} that the intensity distribution from plane-parallel model stellar atmospheres do not vary significantly with respect to fundamental parameters, whereas spherically-symmetric models do.  
In this section, one can conclude that intensity profiles from plane-parallel model atmospheres are insensitive to assumed fundamental parameters, hence the pulsation period.  On the other hand, intensity profiles from spherically-symmetric model atmospheres are more centrally-concentrated  with increasing pulsation period, i.e. the amount of atmospheric extension depends on the pulsation period.  This is important because the projection factor depends on the structure of intensity profiles, hence how the projection factor depends on pulsation period also depends on the assumed geometry of model atmospheres. This also suggests that limb darkening is a more significant effect in measuring Cepheid angular diameters and p-factors than found previously.

\section{Angular diameter corrections}
\cite{Marengo2002} found that limb darkening varied significantly as a function of pulsation phase, by an amount greater than the variation shown in Fig.~\ref{f1} for differences in fundamental parameters.  Limb-darkening corrections, $\theta_{\rm{UD}}/\theta_{\rm{LD}}$, however, do not change \citep{Marengo2003}.  Limb-darkening corrections are the ratio between the uniform disk angular diameter, $\theta_{\rm{UD}}$, which assumes the intensity at any point on the star is equal to the central intensity and the actual limb-darkened angular diameter, $\theta_{\rm{LD}}$ \citep{Davis2000}. 
\cite{Kervella2004} presented interferometric observations of nine Galactic Cepheids to calibrate the surface-brightness relation, however limb-darkened angular diameters presented depended on theoretical limb-darkening corrections from \cite{Claret2000}.  Neither \cite{Claret2000} nor \cite{Marengo2003} considered the role of geometry in predicting limb-darkening relations or limb-darkening corrections.
Because the central intensity of a spherically-symmetric model atmosphere is greater than that for a plane-parallel model for a given stellar flux, then $\theta_{\rm{UD}}$ for a spherical model will be smaller, meaning a smaller limb-darkening correction.  This suggests that \cite{Marengo2002, Marengo2003} overestimated the limb-darkening corrections for Cepheids as well as how much pulsation affects the intensity profile as a function of pulsation phase. 

\begin{figure}[t]
\begin{center}
\includegraphics[width=0.5\textwidth]{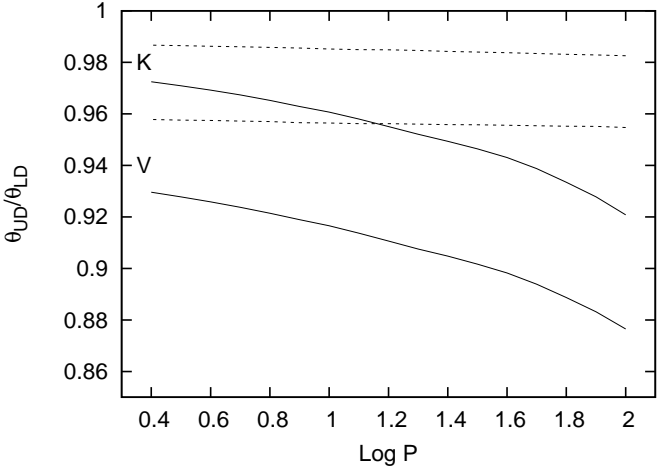}
\end{center}
\caption{Predicted angular diameter corrections $\theta_{\rm{UD}}/\theta_{\rm{LD}}$ as a function of pulsation period computed from spherically-symmetric (solid lines) and plane-parallel (dotted lines) model stellar atmospheres.  In the plot, the $K$-band corrections are greater than the $V$-band corrections.}\label{dia}
\end{figure}

In Fig.~\ref{dia}, we present angular diameter corrections as a function of pulsation computed from plane-parallel and spherically-symmetric model stellar atmospheres for $V$- and $K$-bands. Corrections computed from plane-parallel model atmospheres do not vary by more than $0.3\%$ for all periods, but the spherically- symmetric limb-darkening corrections vary from $k = 0.93$ to $0.88$ in the $V$-band and $k=0.98$ to $0.92$ in the $K$-band.   These corrections vary significantly, both from plane-parallel corrections and as a function of pulsation period.  These results suggest that \cite{Kervella2004} underestimated the angular diameters of the observed Cepheids, and may explain the small but significant differences between surface brightness relations measured by \cite{Kervella2004} and \cite{Fouque1997}.  The result is also consistent with the results of \cite{SasselovK1994} and \cite{Nardetto2004}.

\section{Period - projection factor relation}
Using our predicted intensity profiles, geometric p-factors ($p_0$) are computed using Eq.~\ref{e1},  as functions of pulsation period and wavelength. Projection factors are computed as a function of period as $p = p_0f_{\rm{grad}}f_{o-g}$, where $f_{\rm{grad}}$ is a correction due to velocity gradient within the atmosphere of the Cepheid, and $f_{o-g}$ a correction due to the differential photospheric pulsation velocity between optical and gas layers \citep[see][for more details]{Nardetto2007}.  We assume that $f_{\rm{grad}} = -0.02\log P + 0.99$, and $f_{o-g} = -0.023\log P + 0.979$ from \cite{Nardetto2007}.  We then multiple our Pp relation by $0.95$ to correct for the dependence of the p-factor on the the cross-correlation method. This computation of the p-factor allows us to explore if differences in stellar limb-darkening between the models computed here and by \cite{Nardetto2009} may resolve the differences between theoretical hydrodynamic Pp relations and those derived using the IRSB technique.  If the difference remains then we need to consider other uncertainties and phenomena. 

\begin{figure}[t]
\begin{center}
\includegraphics[width=0.5\textwidth]{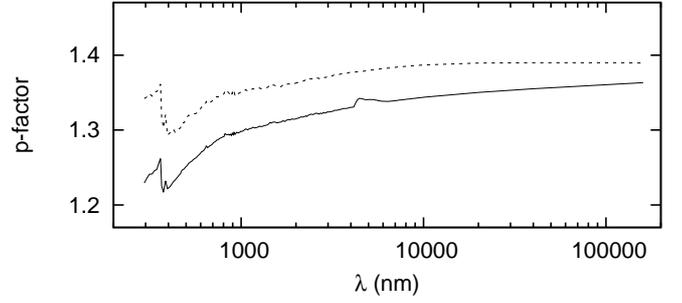}
\end{center}
\caption{The predicted projection factor as a function of wavelength for the $P = 10$~day model atmosphere for the spherically-symmetric (solid line) and plane-parallel (dashed line) geometries.}\label{f4}
\end{figure}

 We start by considering the p-factor as a function of wavelength for the $\log P = 1$ model stellar atmospheres assuming both plane-parallel and spherically-symmetric geometries, shown in Fig.~\ref{f4}.  The p-factor differs significantly as a function of geometry where those from plane-parallel model atmosphere are $3$ - $7\%$ larger.   It also varies as a function of wavelength: for the spherically-symmetric $\log P = 1$ model stellar atmosphere the p-factor varies from about $p = 1.22$ at $\lambda \approx 400~$nm to about $p = 1.37$ as $\lambda \rightarrow \infty$.  This is consistent with the differences between $V$- and $K$-band intensity profiles shown in Fig.~\ref{f1}.  At shorter wavelengths, the intensity profile is more centrally concentrated than intensity profiles in the near infrared, suggesting a smaller p-factor.

\begin{figure}[t]
\begin{center}
\includegraphics[width=0.5\textwidth]{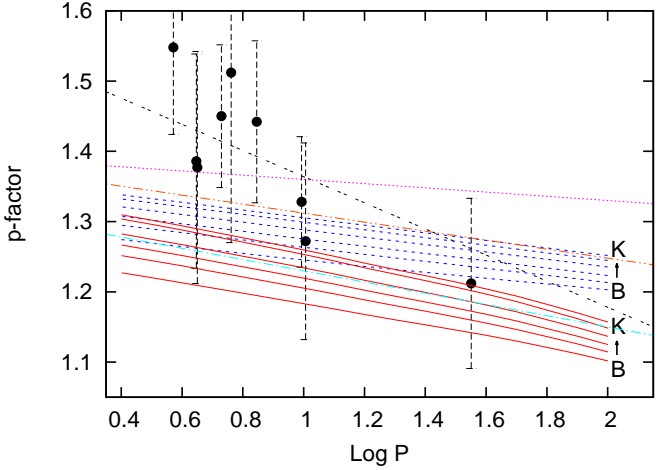}
\end{center}
\caption{The p-factor predicted from plane-parallel and spherically-symmetric model atmospheres as a function of pulsation period.  Projection factors computed from plane-parallel models are shown for the $BVRIH$ and $K$-bands as blue dashed lines, while projection factors from spherically-symmetric models are shown as red solid lines.  For reference the period-projection factor relations are shown from \cite{Hindsley1986} (magenta dotted line), \cite{Nardetto2007}, their Eq.~6 (orange dashed-double dot line), \cite{Nardetto2009} (pale blue dot-dashed line) and \cite{Storm2011b}, (black double-dashed line). The projection factors for Galactic Cepheids determined by \cite{Storm2011a} are also included (black dots). }\label{f5}
\end{figure}

The next step is to compute the p-factor as a function of pulsation period.  This is done for $BVRIH$ and $K$-band intensity profiles for both model geometries, and is shown in Fig.~\ref{f5}. The plane-parallel model geometric p-factors are constant with respect to pulsation period, variation of the p-factor seen in Fig.~\ref{f5} is due to variation of $f_{o-g}(\log P)$ and $f_{grad}(\log P)$. The spherically-symmetric model p-factors decrease as a function of period non-linearly.  In this case, we are comparing geometric p-factors from our hydrostatic plane-parallel and spherically-symmetric model atmospheres to p-factors for four cases.  The first case is the Pp relation from plane-parallel model atmospheres compared to spectra from \cite{Hindsley1986}, but also see \cite{Parsons1972} and \cite{Burki1982}.  The second case is Pp relations from spherical models, Eq.~6 in \cite{Nardetto2007}, while the third case is the \cite{Nardetto2009} relation.  The former relation uses the first moment method, while the latter assumes radial velocities are measured using the cross-correlation method. The fourth case is the LMC+HST period-projection factor relation from \cite{Storm2011a, Storm2011b}, though we could also include results from \cite{Merand2005}, \cite{Groenewegen2007} and \cite{Ngeow2011}.  The LMC+HST Pp relation from \cite{Storm2011a} has a slope measured from LMC Cepheids while the zero-point is measured from Galactic Cepheids. Our results appear to agree with the \cite{Nardetto2009} relation, but not the observed relation, verifying that limb-darkening does not explain the difference between the observed and theoretical relations, hence we must consider different phenomena to explain the difference between the relations from \cite{Nardetto2009} and \cite{Storm2011a}.

The LMC+HST Pp relation is based on observations of lower-metallicity Cepheids, while our models are computed at solar metallicity.  If stellar limb-darkening varies significantly as a function of metallicity, then the Pp relations might be resolved, however, \cite{Nardetto2011} determined from hydrodynamic models that metallicity does not resolve that difference.  We computed another set of model stellar atmospheres following the same period-radius, period-gravity and period-bolometric luminosity relations, but at LMC metallicity.  At lower metallicity, the p-factor is marginally larger than at solar metallicity;  by at most $3\%$ for our $P = 100~$day Cepheid.  Only for the longest-period Cepheid does limb darkening differ significantly.  We note that these models test only differences in metallicity and not differences in pulsation period due to differences in metallicity.  To test that effect, we would require a LMC period-bolometric luminosity relation (the other two relations are believed to be metallicity independent), which is not available.  This suggests metallicity differences between LMC and Galactic Cepheids do not lead to significantly different p-factors and agrees with \cite{Nardetto2011}.
\begin{table}[t]
\caption{Best-fit period-geometric projection factor relations from spherical model atmospheres.}\label{t1}
\begin{center}
\begin{tabular}{lcccc}
\hline
\hline
\multicolumn{5}{c}{Linear: $p_0 = a\log P + b$} \\
\hline
& a & b & $\sigma $& \\
\hline
B&$-0.0366\pm 0.0010$&$1.370\pm0.001$&$0.0020$&\\
V&$-0.0440\pm 0.0015$&$1.402\pm0.002$&$0.0030$&\\
R&$-0.0465\pm 0.0017$&$1.420\pm0.002$&$0.0034$&\\
I&$-0.0486\pm 0.0018$&$1.438\pm 0.002$&$0.0036$&\\
H&$-0.0546\pm0.0021$&$1.465\pm 0.003$&$0.0042$&\\
K&$-0.0525\pm0.0020$&$1.471\pm0.002$&$0.0040$&\\
\hline
\multicolumn{5}{c}{Quadratic: $p_0 = c(\log P)^2 +d$} \\
\hline
& c&d&$\sigma$ &\\
\hline
B &$-0.0150\pm 0.0004$&$1.351\pm0.001$&$0.0019$&\\
V&$-0.0181\pm 0.0003$&$1.379\pm0.001$&$0.0016$&\\
R&$-0.0192\pm0.0003$&$1.396\pm0.001$&$0.0016$&\\
I& $-0.0201\pm0.0003$&$1.414\pm0.001$&$0.0017$&\\
H&$-0.0226\pm0.0003$&$1.438\pm0.001$&$0.0017$&\\
K&$-0.0217\pm0.0003$&$1.444\pm0.001$&$0.0017$&\\
\hline
\multicolumn{5}{c}{Power-Law:$p_0 = x(\log P)^z +y$} \\
\hline
& x&y&z& $\sigma$ \\
\hline
B&$-0.0224\pm0.0016$&$1.357\pm 0.001$&$ 1.51\pm0.08$&$0.0010$\\
V&$-0.0236\pm0.0014$&$1.384\pm0.001$&$ 1.67\pm0.07$&$0.0011$\\
R&$-0.0239\pm0.0014$&$1.400\pm0.001$&$1.72\pm0.07$&$0.0012$\\
I&$-0.0247\pm0.0015$&$1.418\pm0.001$&$1.74\pm 0.07$&$0.0013$\\
H&$-0.0268\pm0.0016$&$1.441\pm0.001$&$1.78\pm0.07$&$0.0014$\\
K&$-0.0262\pm0.0017$&$1.448\pm0.001$&$1.76\pm0.08$&$0.0014$\\
\hline
\end{tabular}
\end{center}
\end{table}

The predicted geometric p-factors calculated from spherically-symmetric models are fit with various relations.  We assume a linear, quadratic, and power-law relations. The best-fit coefficients for each relation are presented in Table~\ref{t1}. The predicted linear geometric relations are consistent with results from \cite{Hindsley1986} as well as \cite{Parsons1972} and \cite{Nardetto2007} (their Eq.~5 for the geometric p-factor).  The quadratic relations do not fit the predicted projection factors much better than a linear model, but the power-law fits best with an exponent ranging from $1.5$ to $1.75$. Thus, geometric p-factors, hence p-factors themselves are non-linear functions of period.

\begin{figure}[t]
\begin{center}
\includegraphics[width=0.5\textwidth]{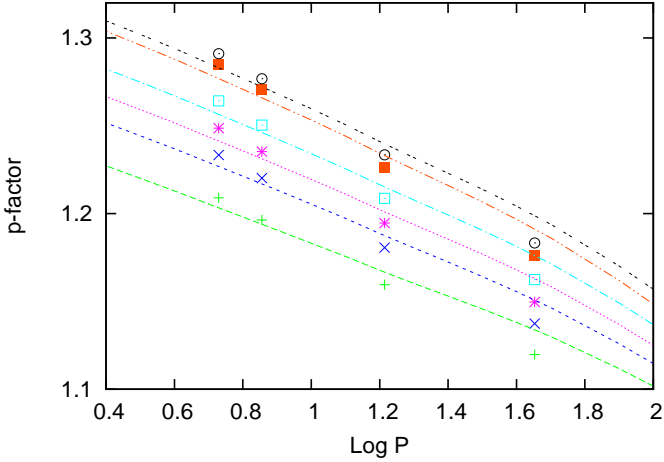}
\end{center}
\caption{Projection factors predicted from  spherically-symmetric model atmospheres representing a sample of Galactic Cepheids as a function of waveband, from $B$ to $K$-band.  In the plot, the colour green represents $B$-band,  blue is $V$, magenta is $R$, light blue is $I$, orange is $H$ and black is $K$-band. }\label{f6}
\end{figure}

\begin{table}[t]
\caption{Parameters for Galactic Cepheids with model stellar atmospheres}\label{t2}
\begin{center}
\begin{tabular}{lccccc}
\hline
\hline
Name & Period &$T_{\rm{eff}}$&$R$&$\log g$ \\
& days & $(K)$&$(R_\odot)$& (cgs)\\
\hline
$\delta$ Cep&5.37&5956&42.6&2.0 \\
$\eta$ Aql &7.18&5772&49.0& 1.8 \\
X Cyg & 16.39& 5283&100.6& 1.15 \\
SV Vul &45.00&  5224&168.7&0.7\\
\hline
\end{tabular}
\end{center}
\end{table}
As a consistency check, we compute model stellar atmospheres for a sample of Galactic Cepheids, listed in Tab.~\ref{t2} with measured gravities, effective temperatures and radii, derived using the IRSB technique from \cite{Storm2011a}. Projection factors for these Cepheids are computed as a function of wavelength, which we compare to our predicted Pp relations in Fig.~\ref{f6}.  The predicted p-factors for each Cepheid are similar to the predicted Pp relations for each waveband,  however the Cepheid p-factor is sensitive to the measured gravity of the Cepheid.  For instance, if we compute a model with the same parameters as SV Vul but with $\log g = 1.15$ then the p-factor increases by about 3-5\%.  The sensitivity of the p-factor on the assumed gravity also grows with decreasing gravity, i.e. increasing pulsation period.  This may help explain some of the difference between the LMC-HST and cross-correlated hydrodynamic Pp relations.

\section{Why do theory and observations disagree?}
There is a clear difference between LMC-HST and cross-correlated hydrodynamic Pp relations, with no obvious resolution. We explore three possibilities: Cepheid mass loss and infrared excess; the effect of Cepheid chromospheres; and possible dispersion of the Pp relation due to the possible range of Cepheid gravities.  The difference between the two types of Pp relations may also be related to the uncertainty of radial velocity measurements  due to line profile asymmetries.  \cite{Ngeow2011} discusses how differences in radial velocity measurements change the determined p-factor for $\delta$ Cephei.  In terms of the Pp relation, these differences will not lead to a systematic shift of the Pp relation but will lead to a greater scatter of the p-factors used to fit a relation.

Recent observations show that Cepheids undergo mass loss \citep{Marengo2010a, Marengo2010b, Barmby2011, Matthews2011, Neilson2011c}.  This mass loss is believed to lead to a far-infrared excess and may be consistent with the K-band excess observed by \cite{Kervella2006,Merand2006,Merand2007}. \cite{Neilson2010} showed that LMC Cepheids appear to undergo mass loss based on measured optical and IR fluxes compiled by \cite{Ngeow2009}.  From this analysis, \cite{Neilson2010} showed that Cepheid mass loss affects the observed $(V-K)_0$ colors, hence the infrared surface brightness method is also affected.  In that case, the IRSB technique predicts a $3-4\%$ larger angular diameter than reality, at least for shorter periods. This leads to the Pp relation that is also $3-4\%$ overestimated.  This difference is not enough to explain the difference between  cross-correlated hydrodynamic and LMC-HST Pp relations, but more detailed observations are needed \citep{Kervella2011}.

Pulsation-driven chromospheric activity is another possibility.  Phase-dependent chromospheric activity leads to limb-brightening near the edge of the stellar disk \citep{Sasselov1994a, Sasselov1994b}.  Consider a Cepheid with a simple limb-darkening relation
\begin{equation}
\frac{I(\mu)}{I(\mu=1)} = 1-\frac{3}{5}(1-\mu),
\end{equation}
from \cite{Getting1935}, which leads to a p-factor of $p=1.41$. If we assume limb-brightening near the edge of the stellar disk $\delta I$ then the geometric p-factor becomes
\begin{equation}
p_{\rm{new}} = \frac{\mathcal{H}}{K}  = \frac{\int (I(\mu)+\delta I(\mu))\mu d\mu}{\int (I(\mu)+\delta I(\mu))\mu^2 d\mu} = \frac{\mathcal{H}_0 + \delta \mathcal{H}}{K_0},
\end{equation}
where we assume that terms $\int \delta I \mu^2 d\mu \approx 0$.  As a result,
\begin{equation}
p_{\rm{new}} = 1.41\left(1 + \frac{\delta \mathcal{H}}{\mathcal{H}_0}\right).
\end{equation}
If  $p_{\rm{new}} = 1.5$ then $\delta \mathcal{H}/\mathcal{H}_0 = 0.064$. This suggests a $6\%$ increase of stellar flux due to a change of intensity near the limb of the stellar disk.  Assuming a limb-brightening relation $\delta I(\mu)/I(\mu=1) = a$, where $a$ is constant when $\mu < 0.05$ and zero elsewhere, then for $a  = 20$, $\delta\mathcal{H}/\mathcal{H}_0 = 0.06$.  It is unlikely that a Cepheid chromosphere can produce the necessary intensities to resolve this difference in the p-factor, but may still contribute. Pulsation-generated chromospheres are believed to be generated by shocks and shock properties depend on pulsation period and effective temperature \citep{Neilson2008a}.  At short period, shocks generated by pulsation are stronger and more frequent than at longer period, furthermore, X-ray and UV observations of short-period Cepheids suggest the presence of very hot $>10^5~K$ plasma in the envelope that may be related to chromospheres \citep{Engle2009}.  This simple test suggests that chromospheric effects may affect the Pp relation, for instance,  \cite{Sabbey1995} did find a value $p = 1.6$ from hydrodynamic models suggesting the possibility of limb-brightening in spectral lines as a function of pulsation phase.

A third possible resolution is related to uncertainties of Cepheid parameters, in particular gravity. 
We showed in the previous section that a change of $\Delta \log g = 0.35$ for SV Vul changes the predicted p-factor by about 3-5\% depending on wavelength.  Cepheids can have a large range of gravity based on the range of known radius from about $30$ - $200~R_\odot$, but also because of their masses.  Cepheid masses are uncertain; this is the Cepheid mass discrepancy \citep{Keller2008} and may also be affected by Cepheid mass loss \citep{Neilson2011a}.  The combined dispersion of masses and radii as a function of pulsation period leads to a significant dispersion in Cepheid limb darkening, hence the Pp relation.  \cite{Nardetto2007} measured the Pp relation from a sample of eight Cepheid models.  The predicted relation may be affected by the small sample that does not account for the dispersion of the Pp relation due to differences in gravity for the same pulsation period.

Unfortunately, none of these three phenomena are obviously sufficient to uniquely resolve the differences between  cross-correlated hydrodynamic and LMC+HST Pp relations, but a combination of the three effects might do so. This analysis also highlights just how sensitive a precision understanding of the p-factor is to the physics describing Cepheids.

\section{Summary}
The Cepheid projection factor is one of the most significant sources of uncertainty for calibrating the Cepheid Leavitt law to the precision necessary to constrain cosmological parameters. This uncertainty is compounded by the fact that various methods for determining the projection factor and its dependence on pulsation period do not agree. This uncertainty is also affected by errors in previously published angular diameter corrections used for interferometric observations.  Plane-parallel model atmospheres predict limb-darkening corrections that are $3$ to $6\%$ larger than spherically-symmetric model atmospheres.  This difference will be explored in greater detail in a future article.

In this work, we computed p-factors from intensity profiles predicted using spherically-symmetric model stellar atmospheres as a function of wavelength. Each model is computed for assumed values of luminosity, radius and gravity, each a function of pulsation period. From the models, the Pp relation is predicted for a number of wavebands and compared to the results from numerous works, but do not agree with observed relations \citep{Storm2011a}.  However, the predicted projection factors are consistent with measured projection factors for Galactic Cepheids with astrometric distances. 

We also explored three possible reasons why there is such significant disagreement between  cross-correlated hydrodynamic and LMC-HST Pp relations.  Cepheid mass loss and infrared excess may affect the observed colors, in turn adding uncertainty to the predicted angular diameters measured using the IRSB technique, or  cross-correlated hydrodynamic models do not account for potential limb-brightening due to pulsation-driven chromospheres.  Conversely, both LMC-HST and cross-correlated hydrodynamic relations are derived by different samples, where the measured p-factors are very sensitive to the gravity of Cepheids, which is in turn, has a significant dispersion as a function of pulsation period.  None of these three hypotheses appear to resolve the differences, but could contribute to a portion of the differences.  This does highlight the sensitivity of the p-factor to atmospheric dynamics and Cepheid properties.

In summary, we conclude that spherically-symmetric model stellar atmospheres provide insight into understanding the projection factor, the Pp relation and the wavelength dependence and dispersion of these relations.  These models also hint at a possible nonlinear Pp relation that reflects the nonlinear nature of Cepheid atmospheric extension as a function of period.  This non-linearity may be more pronounced if we employed a nonlinear bolometric Leavitt Law, similar to the nonlinear optical Leavitt Law \citep{Kanbur2004}.

\bibliographystyle{aa} 

\bibliography{pfact}
\end{document}